\documentclass[universe,article,accept,pdftex,moreauthors]{Definitions/mdpi} 
\usepackage{environ}
\NewEnviron{myequation1}{%
\begin{equation}
\scalebox{0.88}{$\BODY$}
\end{equation}                                
}
\firstpage{1} 
\pubvolume{1}
\issuenum{1}
\articlenumber{0}
\pubyear{2024}
\copyrightyear{2024}
\datereceived{3 July 2024} 
\daterevised{24 July 2024 } 
\dateaccepted{25 July 2024 } 
\datepublished{ } 
\hreflink{https://doi.org/} 

\Title{Low-Energy Cosmic Rays and Associated MeV Gamma-Ray Emissions in the Protoplanetary System}

\TitleCitation{Low-Energy Cosmic Rays and Associated MeV Gamma-Ray Emissions in the Protoplanetary System}

\Author{Xulei Sun $^{1}$\orcidA{}, Shuying Zheng $^{2}$, Zhaodong Shi $^{3,}$*\orcidB{}, Bing Liu $^{3}$\orcidC{}, and Ruizhi Yang $^{3}$\orcidD{}}

\AuthorNames{Xulei Sun, Shuying Zheng, Zhaodong Shi, Bing Liu and Ruizhi Yang}

\AuthorCitation{Sun, X.; Zheng, S.; Shi, Z.; Liu, B.; Yang, R.}

\address{%
$^{1}$ \quad School of the Gifted Young, University of Science and Technology of China, Hefei 230026, China; sunxulei@mail.ustc.edu.cn (X.S.)\\
$^{2}$ \quad Department of Modern Physics, School of Physical Sciences, University of Science and Technology of China, Hefei 230026, China; zsy5213@mail.ustc.edu.cn (S.Z.)\\
$^{3}$ \quad Department of Astronomy, School of Physical Sciences, University of Science and Technology of China, Hefei 230026, China; lbing@ustc.edu.cn (B.L.); yangrz@ustc.edu.cn (R.Y.)}

\corres{Correspondence: shizd@ustc.edu.cn (Z.S.)}

\abstract{Low-energy cosmic rays (LECRs) play a crucial role in the formation of planetary systems, and detecting and reconstructing the properties of early LECRs is essential for understanding the mechanisms of planetary system formation. Given that LECRs interact with the surrounding medium to produce nuclear de-excitation line emissions, which are gamma-ray emissions with energy mainly within 0.1--10 MeV and are unaffected by stellar wind modulation, these emissions can accurately reflect the properties of LECRs. This study introduces an innovative method for using gamma-ray emissions to infer LECR properties. We employed the Parker transport equation to simulate the propagation and spectral evolution of LECRs in a protoplanetary disk and calculated the characteristic gamma-ray emissions resulting from interactions between LECRs and disk material. These gamma-ray emissions encapsulate the spectral information of LECRs, providing a powerful tool to reconstruct the cosmic ray environment at that time. This method, supported by further theoretical developments and observations, will fundamentally enhance our understanding of the impact of CRs on the origin and evolution of planetary systems and address significant scientific questions regarding the cosmic ray environment at the origin of life.}


\begin{document}


\section{Introduction}
Cosmic rays (CRs) are high-energy particles from outer space, predominantly composed of protons and atomic nuclei \cite{ref-becker2020}. These particles exhibit a wide energy range, from tens of MeV to over $10^{20}$ eV. Over the past two decades, there have been notable advancements in the measurement of cosmic rays (CRs), with unprecedented precision~\cite{AMS2021, IceCube2012, IceCube2013, IceTop2013a, IceTop2013b}, and to some extent, we have also gained insight into their distribution over the Galaxy \cite{aharonian2020}. On the other hand, low-energy CRs (LECRs), which typically have energies ranging from a few MeV to several GeV per nucleon, remain a subject of ongoing investigation. Despite the availability of energy spectra in the local interstellar medium of LECRs thanks to Voyager 1 and 2 \cite{cummings2016, stone2019}, our comprehension of LECRs remains limited due to the solar modulation when they come into the heliosphere and to the lack of powerful observational methods for probing their distribution in the interstellar medium (ISM). As a significant component of the ISM, LECRs have profound impacts on the ionization, heating, and chemical evolution of the ISM due to their high energy-loss rates within the ISM \cite{ref-grenier2015, ref-gabici2022, ref-padovani2020}. Consequently, it is imperative to investigate the effects of LECRs on the ISM, and stellar and planetary systems, etc.

Protoplanetary disks are dense gas and dust disks that form around young stars during the collapse of molecular clouds \cite{ref-williams2011}. These disks, with radii extending up to 1000 AU, are the birthplaces of planetary systems. LECRs significantly impact protoplanetary disks when interacting with the gases and dust in the disk. LECRs can effectively ionize the gas in the protoplanetary disk, altering its ionization level and electrical conductivity, which in turn affects the coupling efficiency of magnetic fields and the excitation of turbulence~\mbox{\cite{ref-cleeves2013, ref-cleeves2014, ref-padovani2018}}, changing the dynamic structure and evolutionary path of the disk. Additionally, LECRs provide an extra heating source for the disk, influencing its temperature distribution and geometric shape \cite{ref-glassgold2012}. Changes in the disk temperature can affect the evolution of its chemical composition and dust properties. Furthermore, LECRs can induce various photochemical reactions in molecules and atoms, altering the chemical composition of the disk \cite{ref-oberg2021, ref-walsh2010}. These chemical changes might impact the composition of planetary atmospheres and the synthesis of prebiotic molecules. LECRs interacting with surrounding material produce gamma-ray and X-ray radiation, modifying the radiation environment of the protoplanetary disk. This high-energy radiation can affect the ionization and chemical state of the disk's material. Moreover, LECRs can interact with dust particles, changing their charge, size, and composition \cite{ref-oberg2023}. This affects the aggregation of dust and the formation process of planets.

Studying the propagation and spectral evolution of LECRs in protoplanetary disks is thus crucial for understanding the formation and early evolution of planetary systems. However, due to the slow propagation and energy loss of sub-relativistic particles, LECRs are concentrated near their acceleration sites, making them difficult to detect directly. 
They are often studied through indirect measurements of the molecular ionization rate and non-thermal radiation, such as the 6.4 KeV Fe K$\alpha$ line emission. Unfortunately, these methods cannot rule out the influence of CR electrons and UV photons, leading to considerable uncertainties in the results. In the protoplanetary disk environment, LECRs interacting with gas and dust may produce a significant amount of nuclear de-excitation line emission with energy mainly within the range of 0.1--10 MeV. The gamma-ray line emission emitted via nuclear de-excitation following the collisions of LECR nuclei with the ambient gases have strong penetration capability and are unaffected by magnetic fields and stellar winds, thus providing a very effective probe to investigate the acceleration and transport processes of LECRs and their composition \cite{ref-ramaty1979a, ref-ramaty1979b, ref-kozlovsky2002} and offering important observational evidence for exploring the physical processes and chemical evolution within the protoplanetary disk.

The nuclear de-excitation gamma-ray lines generated during solar flares have been extensively investigated observationally \cite{ref-chupp1973, ref-cliver1989, ref-vilmer2011} and theoretically \cite{ref-ramaty1975, ref-murphy2009}. The nuclear de-excitation gamma-ray line emission has also been proposed to be a promising probe to study the LECRs in the interstellar medium \cite{ref-mezhoud2013} and in the vicinity of their sources \cite{ref-liu2021} such as in the supernova remnants \cite{ref-liu2023} and in the Galactic center region \cite{ref-dogiel2009}. 
The present study investigates the LECRs in protoplanetary disks through the simulation of LECR propagation in such disks and the calculation of corresponding nuclear de-excitation line emission. It may shed light on the high-energy particle environment during the formation processes of protoplanetary systems from future observations.




The paper is structured as follows: Section \ref{sec:lecr} simulates the propagation and evolution of LECRs in protoplanetary disks. Section \ref{sec:gamma} calculates and examines the gamma-ray radiation characteristics resulting from LECR interactions. Section \ref{sec:disc} discusses potential improvements to the model. Section \ref{sec:summ} provides the summary and outlook.

\section{Calculation of LECR Spectrum} \label{sec:lecr}
This section introduces the numerical methods employed to solve the LECR spectrum, specifically the numerical algorithm and code implementation details for solving the Parker transport equation. The evolution of LECRs is crucial for understanding the chemical processes in protoplanetary disks and the ionization environment of early Earth. Therefore, accurate calculation of the LECR spectrum is essential.

\subsection{Theoretical Basis}
\subsubsection{Parker Transport Equation}
CRs from outside the astrosphere are modulated by the stellar wind during their transport in a solar-type stellar system. The transport of CR particles in the interplanetary medium can be described by the Parker equation \cite{ref-parker1965}:
\begin{equation}
\frac{\partial f}{\partial t} + \mathbf{v}\cdot\nabla f - \nabla\cdot (\kappa\nabla f) - \frac{1}{3}(\nabla\cdot\mathbf{v})p\frac{\partial f}{\partial p} = 0,
\end{equation}
where \(f(\mathbf{r}, p, t)\) is the phase space distribution function of CR particles, \(\mathbf{v}\) is the stellar wind velocity field, \(\kappa\) is the diffusion coefficient, and \(p\) is the momentum of CR particles (protons are considered here). The last three terms on the left side of the equation represent spatial advection, spatial diffusion, and adiabatic energy loss, respectively.

While a complete description of the protoplanetary disk and wind system is beyond the scope of the present work, we assume that the stellar wind is spherically symmetric~\cite{ref-rodgers2020} as the lowest order approximation to the system; then, the Parker equation reads:
\begin{equation}
\frac{\partial f}{\partial t} + v_r\frac{\partial f}{\partial r} - \frac{1}{r^2}\frac{\partial}{\partial r}\left(r^2 \kappa \frac{\partial f}{\partial r}\right) - \frac{1}{3}\frac{1}{r^2}\frac{\partial (r^2 v_r)}{\partial r}p\frac{\partial f}{\partial p} = 0,
\label{parker}
\end{equation}
where $r$ is the radial distance from the central star, $v_r$ is the radial velocity of stellar wind, and $p$ is the momentum of LECRs. We will employ numerical methods to solve this equation and obtain the temporal and spatial evolution as well as the spectral distribution of LECRs in the protoplanetary disk.

\subsubsection{Stellar Wind Velocity}
The radial velocity of the stellar wind, \(v_r\), tends to increase with the stellar rotation rate \(\Omega\) \cite{ref-rodgers2020}. At a distance of 1 AU from the star, \(v_r\) is adapted in this study from the fitting formula in \cite{ref-carolan2019}:
\begin{equation}
  \log _{10}v_r = a\Bigg (\frac{\Omega }{\Omega _\odot }\Bigg)^b + c\Bigg (\frac{\Omega }{\Omega _\odot }\Bigg)^d + e\Bigg (\frac{\Omega }{\Omega _\odot }\Bigg)^f+g,
\end{equation}
where the coefficients \(a\sim f\) are given in \cite{ref-carolan2019}. For \( \Omega < 1.4 \Omega_\odot \), an additional term \mbox{\( g = -\log_{10}1.12 \)} is included for a better fit to the data from \cite{ref-rodgers2020}; for \( \Omega \ge 1.4 \Omega_\odot \), \( g \) is set to 0.

\subsubsection{Diffusion Coefficient}
The diffusion coefficient \(\kappa\) is a critical parameter in the Parker equation, describing the scattering and diffusion process of CR particles in a turbulent magnetic field. For LECRs, due to their small gyroradius, they are more likely to be carried and scattered by stellar winds, resulting in a relatively low diffusion coefficient. This study uses a widely accepted semi-empirical model \cite{ref-rodgers2020,ref-jokipii1973,ref-schlickeiser1989}:
\begin{equation}
  \frac{\kappa (r,p,\Omega)}{\beta c} =\eta _0 \left(\frac{p}{p_0}\right)^{1-\gamma }r_\mathrm{L},
\end{equation}
where \(\beta = v/c\) is the relativistic factor, \(\Omega\) is the stellar rotation angular velocity, \(\eta_0 = \left(\frac{B}{\delta B} \right)^2\) represents the level of turbulence in the magnetic field, \(B\) is associated with the energy density of the large-scale magnetic field, and \(\delta B\) is related to the total energy density in the small-scale magnetic turbulence mode. \(p_0 = 3\, \mathrm{GeV}/c\), \(\gamma\) is related to the turbulence power spectrum and is commonly set to \(\gamma = 1\) based on solar wind modulation observations \cite{ref-cohen2012}, and \(r_\mathrm{L} = p/[eB(r,\Omega)]\) is the Larmor radius of a proton. This model effectively captures the behavior of the diffusion coefficient in the inner interplanetary space.

\subsubsection{Boundary Conditions}
To solve the Parker equation, appropriate boundary conditions are required. In the momentum space \(p\) direction, due to the negligible contribution of extremely high-energy CRs to the total spectrum, boundary condition $f(r, p=p_\mathrm{max}, t)$ = 0 is adopted, where $p_\mathrm{max}$ is the maximal momentum in our numerical scheme. 

For the spatial coordinate \(r\), a reflective boundary condition is adopted at the inner boundary:
\begin{equation}
f(r=r_\odot, p, t) = f(r=r_\odot + \Delta r, p, t).
\end{equation}

This implies that CR particles near the stellar surface are reflected to outer space without being lost to the stellar surface.

At the outer boundary in the \(r\) direction, the local interstellar spectrum (LIS) obtained from observations is used as the boundary condition:
\begin{equation}
f(r=r_{\text{LIS}}, p, t) = f_{\text{LIS}}(p).
\end{equation}

The LIS refers to the CR spectrum outside the heliosphere, unaffected by stellar wind modulation. Typically, the outer boundary \(r_{\text{LIS}}\) is set at over 100 AU, far from the star to avoid stellar wind influence. In this study, the heliosphere radius is used as \( r_{\text{LIS}} \) \cite{ref-rodgers2020}:
\begin{equation}
  r_{\text{LIS}}(\Omega) = R_\mathrm{h}(\Omega) = R_\mathrm{h}(\Omega_\odot) \sqrt{\frac{{\dot{M}(\Omega)}v(\Omega)}{{\dot{M}(\Omega_\odot)}v(\Omega_\odot)}},
  \label{equ:heliosphere_radius}
\end{equation}
where \( R_\mathrm{h}(\Omega_\odot) = 122 \) AU is the current heliosphere radius; \( v \) is the radial velocity of the stellar wind at 1 AU; and \( \dot{M} \) is the mass loss rate of the star, with the calculation formula provided in \cite{ref-carolan2019}.

Measurements by Voyager 1 and 2 provide important observational data for the LIS, which can be fitted by the following differential intensity spectrum \cite{ref-vos2015}:
\begin{equation}
  j_\mathrm{LIS}(T) = 2.70\frac{T^{1.12}}{\beta ^2}\left(\frac{T+0.67}{1.67} \right)^{-3.93} \mathrm{m^{-2}}\, \mathrm{s^{-1}}\, \mathrm{sr^{-1}}\, \mathrm{MeV^{-1}},
\end{equation}
where \(j\) is the differential intensity of CRs, representing the number of particles passing through a unit area, per unit time, per unit solid angle, and per unit energy interval; \(T\) is the kinetic energy of CRs in GeV; and \(\beta = v/c\) is the ratio of particle speed to the speed of light. The differential intensity \(j(T) = p^2 f(p)\), where $f(p)$ is the phase space distribution function of CRs.

\subsection{Numerical Methods}
To numerically solve the Parker transport equation, it is necessary to discretize the three independent variables: time (\(t\)), space (\(r\)), and momentum (\(p\)).

Explicit difference schemes for diffusion have to take a sufficiently small timestep $\Delta t$ as required by the Courant–Friedrichs–Lewy condition, thus that it is time-consuming generally. Hence, the semi-implicit Crank--Nicolson scheme \cite{ref-crank1947} for diffusion is adopted (see Appendix \ref{AppA} for further details), while the explicit upwind schemes for advection and adiabatic loss are used. In addition, the operator splitting method is adopted \cite{ref-nr1992}. 



To further improve computational efficiency, we utilize the Numba\endnote{\url{https://numba.pydata.org/} (accessed on 23 November 2023)} \cite{ref-numba2015} library to compile critical calculation steps in Python code into efficient machine code, significantly reducing runtime.

\subsection{Results}
\begin{figure}[H]
  \begin{adjustwidth}{-\extralength}{0cm}
    \centering
      \begin{minipage}[b]{0.65\columnwidth}
        \includegraphics[width=\textwidth,trim=0cm 0cm 0cm 1.5cm,clip]{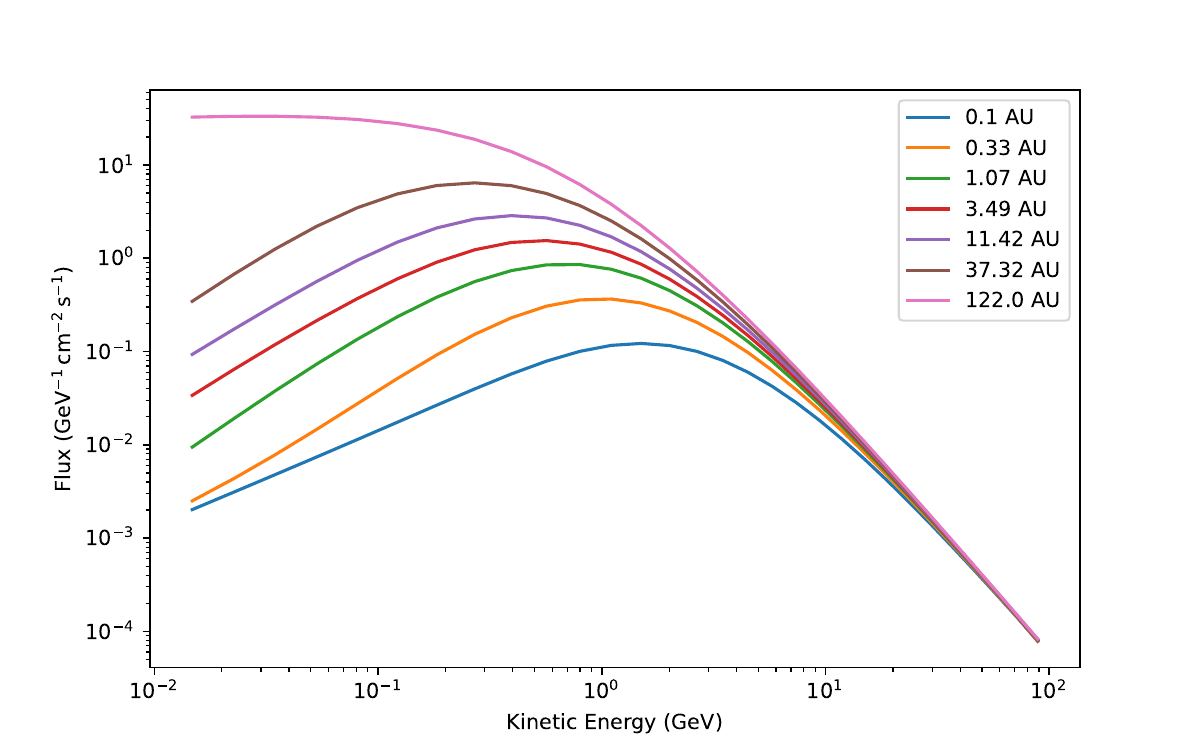}
      \end{minipage}
      \begin{minipage}[b]{0.65\columnwidth}
        \includegraphics[width=\textwidth,trim=0cm 0cm 0cm 1.5cm,clip]{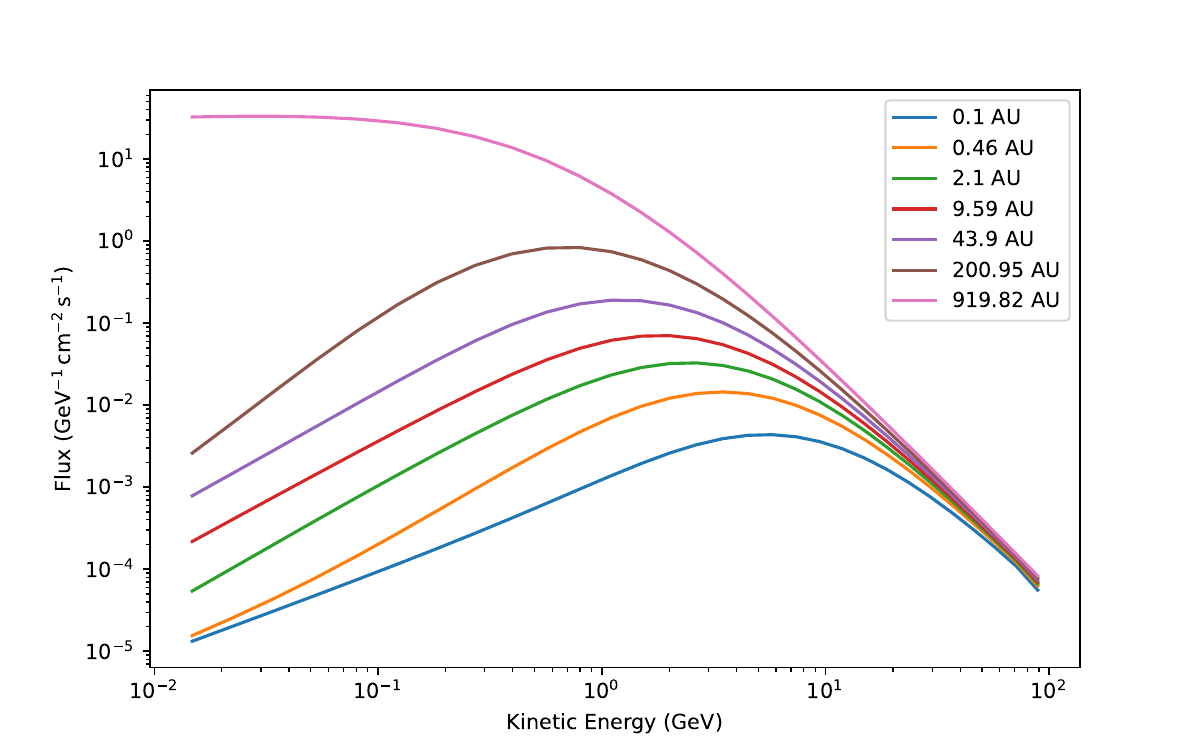}
      \end{minipage}
  \end{adjustwidth}
  \caption{The flux of CR particles (protons are considered here) under the modulation of the stellar wind as a function of kinetic energy at different distances from the star. \textbf{Left panel}: \(\Omega=1.0\,\Omega_\odot\), corresponding to \(r_{\text{LIS}}=122.00\) AU. \textbf{Right panel}: \(\Omega = 2.1 \, \Omega_\odot\), corresponding to \(r_{\text{LIS}}=919.82\) AU.}
  \label{fig:cosmic_spectra}

\end{figure}

By solving the Parker equation with the aforementioned algorithm and parameter settings, we obtain the temporal and spatial evolution of CRs. Figure \ref{fig:cosmic_spectra} presents the momentum spectra of CR particles at different distances for the stellar rotation rate \(\Omega = 1.0 \, \Omega_\odot\) (\(\Omega_\odot\) is the current solar rotation rate) and \(\Omega = 2.1 \, \Omega_\odot\) (it is estimated that the Sun's age was $t = 1\,\mathrm{Gyr}$ when life began on Earth, corresponding to $\Omega\sim 2.1\,\Omega_\odot$).

The figure shows a significant decrease in the flux of LECRs as the distance \(r\) decreases, especially in the low-energy region where the flux is significantly attenuated in the inner interplanetary space. Additionally, the peak flux shifts to higher energies as \(r\) decreases, consistent with the expected effects of diffusion and stellar magnetic field attenuation. The spectrum in the inner interplanetary space shows a steep drop at low energies, while the spectrum in the outer interplanetary space remains relatively flat. This matches our estimates of the protoplanetary disk morphology, where the inner interplanetary space exhibits significant modulation and shielding effects, while the outer interplanetary space is less affected by stellar wind.

Furthermore, compared to the case with $\Omega=1.0\,\Omega_\odot$, the LECR flux at the same location with $\Omega=2.1\,\Omega_\odot$ is smaller. This is understandable—young stars have stronger stellar winds, which inhibit the entry of LECRs.

The shape of the calculated spectrum curves agrees with results from previous related studies \cite{ref-rodgers2020}, validating the reliability of the Parker transport model and the numerical algorithm used in this study. In conclusion, our method can accurately simulate the transport and diffusion behavior of LECRs in the environment of a protoplanetary disk.

\section{Calculation of Nuclear De-Excitation Line Emission} 
\label{sec:gamma}

LECR nuclei interacting with protoplanetary disk material can trigger a series of nuclear reactions, exciting atomic nuclei to high-energy states, which subsequently de-excite by emitting gamma rays in the 0.1--10 MeV range. These gamma rays mainly originate from two processes: collisions between protons and helium nuclei in LECRs with heavy elements in the environmental gas, and collisions between heavy nuclei in LECRs with hydrogen and helium atoms in the environment \cite{ref-ramaty1979b}.

Narrow lines are produced when CR protons and $\alpha$ particles interact with ambient nuclei heavier than helium (such as C, N, O). These interactions result in direct reactions, producing gamma rays with concentrated energy and narrow line widths. This narrowness is due to the relatively low recoil velocity of the heavy nuclei, leading to minimal Doppler shift of the photons. These narrow lines are important features for detecting LECRs.

Broad lines occur when heavy nuclei in CRs (such as C, N, O) are excited by the medium's hydrogen and helium, subsequently de-exciting to produce gamma rays with broader line widths (fractional width of about 20\%). This broader width is because the heavy nuclei retain most of their initial velocity, resulting in significant Doppler shifts.

In this section, we compute the gamma-ray radiation in the protoplanetary disk based on the LECR energy spectrum distribution obtained in Section \ref{sec:lecr}, combined with models of stellar atmospheres and CR elemental abundances, laying the necessary foundation for subsequent discussions.

\subsection{Calculation Method}
\textls[-25]{Following the method developed by~\mbox{\citet{ref-ramaty1979b,ref-murphy2009,ref-mezhoud2013}}}, we applied the same procedure as described in Section 3.1 of \citet{ref-liu2021}. For the nuclear reaction cross sections, we use the TALYS code (version 1.96)\endnote{\url{https://www-nds.iaea.org/talys/} (accessed on 11 April 2024)} \cite{ref-koning2007, ref-koning2023} for theoretical simulations as a complement to the experimental data complied by \citet{ref-murphy2009}. Additionally, the relative abundances of different elements in the protoplanetary disk significantly affect the shape of the gamma-ray radiation spectrum. This study uses stellar atmospheric elemental abundances and CR elemental abundances models from \cite{ref-ramaty1979b}, as shown in \mbox{Table~\ref{tab:element_abundance}.} By multiplying the contribution of each element by its corresponding abundance and summing the weighted contributions of all elements, the total gamma-ray emissivity spectra per hydrogen atom are obtained. Finally, by integrating (in practice, performing a discrete sum in the code) the gamma-ray emissivity throughout the entire protoplanetary disk (from $r=r_\odot \sim 0$ to $r=r_{\text{LIS}}$), and assuming a hydrogen atom number density of 1 cm\(^{-3}\) in the disk, we can calculate the observed gamma-ray flux assuming the star is 1 kpc away from Earth.

\begin{table}[H]
  \centering
  \caption{Elemental abundances relative to hydrogen (i.e., protons) throughout the stellar atmospheres and CRs, taken from \cite{ref-ramaty1979b}.}
  \label{tab:element_abundance}
  \begin{tabularx}{\textwidth}{CCC}
    \toprule
    \textbf{Element} & \textbf{Stellar Atmosphere} & \textbf{CRs} \\
    \midrule
    H & 1 & 1 \\
    He & $10^{-1}$ & $10^{-1}$ \\
    C & $4.2 \times 10^{-4}$ & $3.8 \times 10^{-3}$ \\
    N & $8.7 \times 10^{-5}$ & $5.6 \times 10^{-4}$ \\
    O & $6.9 \times 10^{-4}$ & $4.3 \times 10^{-3}$ \\
    Ne & $3.7 \times 10^{-5}$ & $5.8 \times 10^{-4}$ \\
    Mg & $4.0 \times 10^{-5}$ & $9.2 \times 10^{-4}$ \\
    Al & $3.3 \times 10^{-6}$ & $5.9 \times 10^{-5}$ \\
    Si & $4.5 \times 10^{-5}$ & $8.1 \times 10^{-4}$ \\
    S & $1.6 \times 10^{-5}$ & $1.2 \times 10^{-4}$ \\
    Ca & $2.2 \times 10^{-6}$ & $9.0 \times 10^{-5}$ \\
    Fe & $3.2 \times 10^{-5}$ & $8.5 \times 10^{-4}$ \\
    \bottomrule
  \end{tabularx}
\end{table}



\subsection{Results}
\begin{figure}[htbp]
  \begin{adjustwidth}{-\extralength}{0cm}
    \centering
      \begin{minipage}[b]{0.65\columnwidth}
        \includegraphics[width=\textwidth,trim=0cm 0cm 0cm 1.5cm,clip]{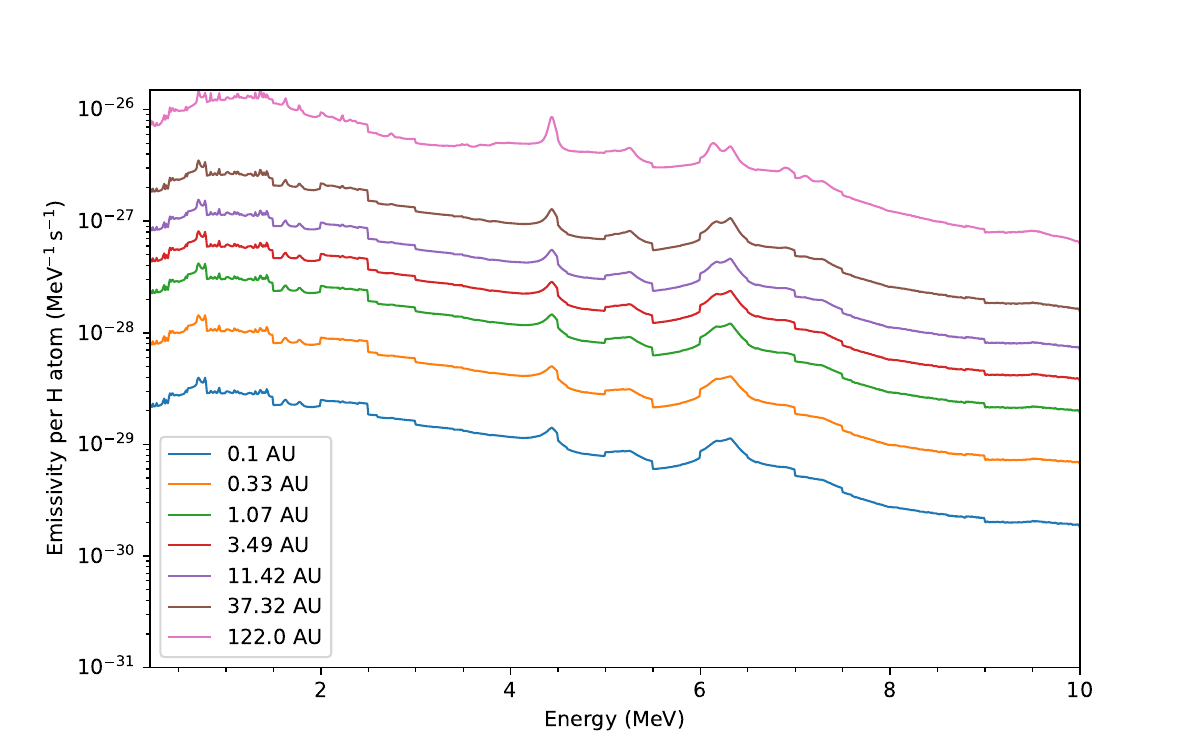}
      \end{minipage}
      \begin{minipage}[b]{0.65\columnwidth}
        \includegraphics[width=\textwidth,trim=0cm 0cm 0cm 1.5cm,clip]{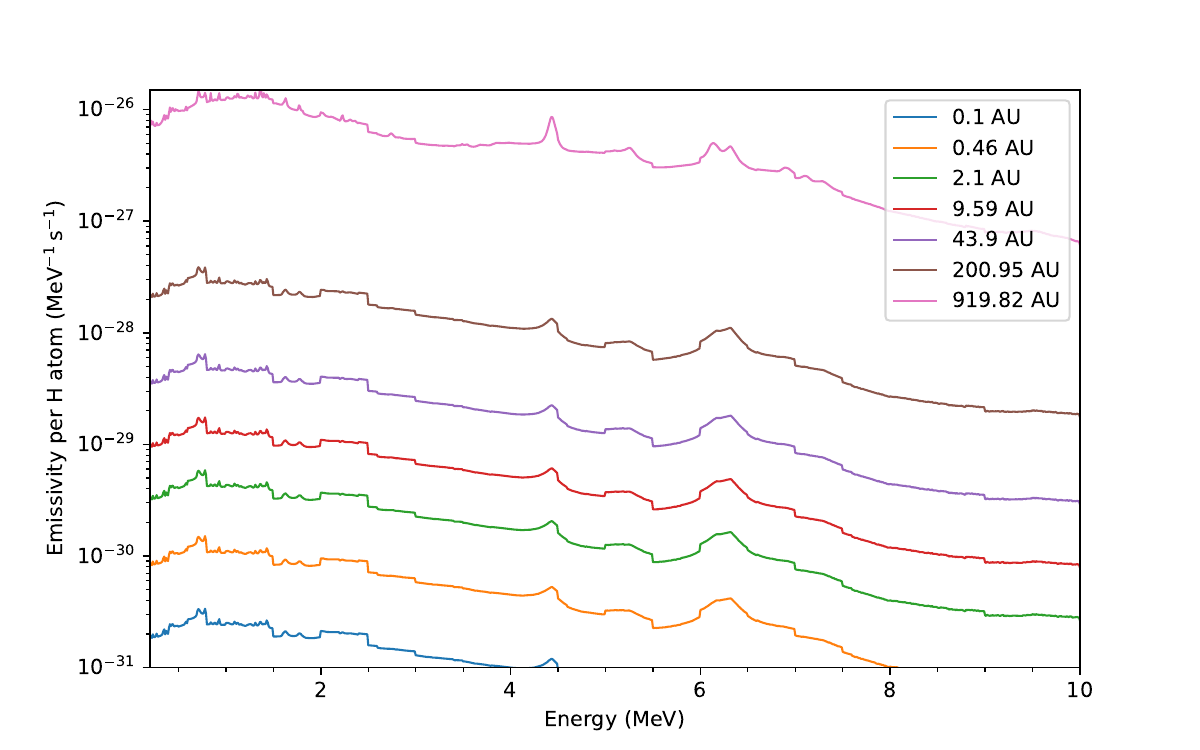}
      \end{minipage}
  \end{adjustwidth}
  \caption{The emissivity of MeV gamma-ray emission per hydrogen atom caused by CRs as a function of energy at different distances from the star. \textbf{Left panel}: \(\Omega=1.0\,\Omega_\odot\). \textbf{Right panel}: \(\Omega = 2.1 \, \Omega_\odot\).}
  \label{fig:gamma_emissivity_spectra}
\end{figure}

Figure \ref{fig:gamma_emissivity_spectra} shows the calculated gamma-ray emissivity at different distances to the center star. Because LECRs are significantly attenuated by the stellar wind in the inner planetary space, the corresponding gamma-ray emissivity decreases as the distance $r$ decreases.

The spectra display some distinct lines, such as the 4.44 MeV line emission mainly resulting from the de-excitation of $^{12}$C and the 6.13 MeV line emission, mainly produced via the de-excitation of $^{16}$O. These spectral lines indicate significant nuclear reactions between LECRs and the protoplanetary disk material.
The gamma-ray emissivity for the case with $\Omega=2.1\,\Omega_\odot$ is lower compared to $\Omega=1.0\,\Omega_\odot$ at the same location. This difference reflects the lower CR flux at that time. Nevertheless, the positions of the spectral lines do not change since the elemental abundance in the protoplanetary disk atmosphere remains the same.

\begin{figure}[H]
  \includegraphics[width=0.8\textwidth,trim=0cm 0cm 0cm 1cm,clip]{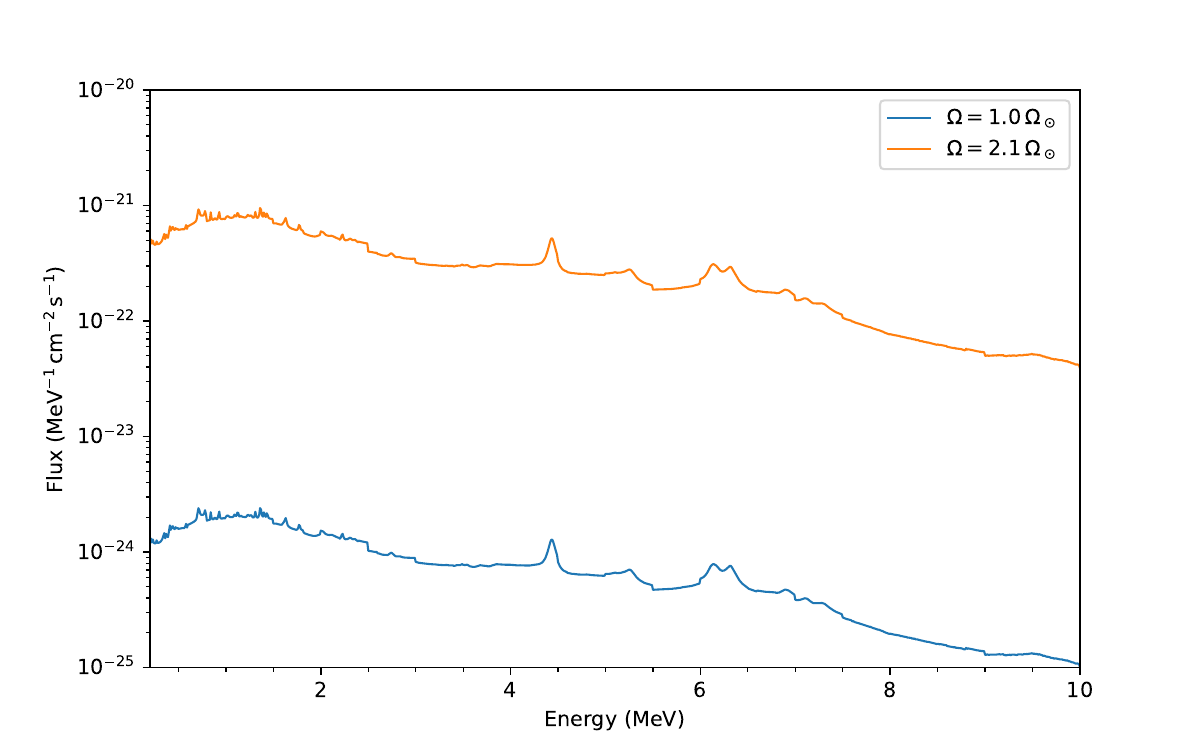}
  \caption{The flux of MeV gamma-ray emission as a function of energy observed at a distance of 1~kpc from the star.}
  \label{fig:gamma_flux_spectra}
\end{figure}

It is challenging to resolve the nuclear de-excitation line emission at different radii, even with the future MeV gamma-ray telescopes, because the angular radius is approximately 1 arcsec, assuming that the radius of the protoplanetary system is about 1000 AU at a distance from Earth of about 1 kpc. Therefore, it is more realistic to calculate the total nuclear de-excitation line emission from the protoplanetary disks. The total gamma-ray line flux is obtained by integrating the emissivity per hydrogen atom multiplied by the gas density over the volume of the protoplanetary system and then dividing by the squared distance. Figure \ref{fig:gamma_flux_spectra} presents the theoretical calculation of the gamma-ray flux spectra supposing the observer is located at a distance of 1 kpc from the star. It is evident that the general trend in the emission spectra is maintained. Furthermore, some of the original spectral lines are preserved, notably the 4.44 MeV de-excitation line of $^{12}$C and the 6.13 MeV de-excitation line of $^{16}$O. Thus, it can reflect the interactions of CRs and elements in the stellar atmosphere. In addition, the gamma-ray flux at $\Omega=2.1\,\Omega_\odot$ is higher, which is a natural consequence of the larger heliosphere radius (expressed in Equation \eqref{equ:heliosphere_radius}) at that time. However, the estimated line fluxes from the protoplanetary system are far below the sensitivity ($\sim 10^{6}\,{\rm ph\,cm^{-2}\,s^{-1}}$) of the next-generation MeV telescopes.

\section{Discussion} \label{sec:disc}

In this article, we calculated the spectra of LECRs in the protoplanetary disk in Section~\ref{sec:lecr} by using Parker's equation \eqref{parker} under the assumption of a spherically symmetric wind, and the resulting nuclear de-excitation gamma-ray line emission spectra in Section \ref{sec:gamma} of LECRs interacting with the protoplanetary disk gases assuming a uniform distribution of material in the disk. These results highlight how the interaction of LECRs with different elements produces a characteristic gamma-ray emission spectrum, consisting of both narrow and broad lines, as well as the unresolved-line continuum (as described in Section 4 of \citet{ref-murphy2009}). Therefore, the structure and intensity of the gamma-ray emission spectrum provide valuable information about the LECRs themselves. Through detailed analysis of the gamma-ray spectrum, the composition, energy, and spatial distribution of LECRs at that time can be coherently inferred, which is crucial for understanding the radiation and ionization environment of early planetary systems.

On the other hand, the protoplanetary disk is a much more complex system than that considered in the present study, and it can be very important to consider a more realistic model for making comparisons with future observations. Hence, we discuss several important points that can further improve our present model below.

It is important to consider the physical and chemical structure and material distribution of the protoplanetary disk. This study adopts a simplified spherical symmetric model, assuming a uniform distribution of material in the protoplanetary disk. The assumption of azimuthal symmetry means that any short-term modulation effects shorter than the stellar rotation period are ignored, but the actual protoplanetary disk may have a flattened disk structure with significantly uneven material distribution perpendicular to the disk plane. The geometric structure will affect the propagation of stellar wind and LECRs in the protoplanetary disk. For example, due to the higher material density in the disk plane, CRs are more likely to undergo scattering and energy loss in this region. Observations also reveal many structural details, such as planetary rings, gaps, and spiral arms \cite{ref-andrews2020}. Therefore, it is necessary to establish a three-dimensional protoplanetary disk model that is closer to the actual situation and study the effects of geometric and material distribution through numerical simulations, such that we can infer more information about the physical and chemical conditions of the protoplanetary disks from gamma-ray observations.
  
Magnetic fields play a significant role in the dynamics and evolution of protoplanetary disks and stellar winds \cite{ref-wardle2007}. Therefore, it is crucial to study the effects of magnetic fields on the transport of LECRs in the protoplanetary disks \cite{ref-cleeves2013, ref-fujii2022}. The current model assumes a uniform diffusion coefficient, ignoring the impact of magnetic field evolution. To address this, it is necessary to coherently simulate the dynamic evolution of the magnetic field in the protoplanetary disk environment and couple it with the calculation of LECR transport. This requires further development of relevant theoretical models and numerical algorithms.


The turbulence of protoplanetary disks \cite{ref-hughes2011, ref-rosotti2023} is highly important for the transport of LECRs. The parameter $\eta_0$ describes the turbulence level in the stellar wind. A higher value means that CRs propagate further before scattering. For example, young stars are believed to produce more coronal mass ejections because they are observed to have higher flare rates, which may lead to stronger turbulent components in the magnetic fields of planetary systems. This study uses $\eta_0 = 1$, representing the turbulence limit of CRs in the stellar wind, which may not accurately approximate the situation for stars. Assuming a constant $\eta_0$ means that the diffusion coefficient decreases with increasing magnetic field strength and therefore also decreases with increasing stellar rotation rate. The value of $\gamma$ is related to the power of turbulence. This study uses $\gamma=1$ because it matches observations of solar wind modulation; however, this approximation may not be accurate for other stars.

Currently, the LECR transport model and the gamma-ray radiation calculations are conducted separately. To obtain coherent results, the two need to be tightly integrated. This involves simultaneously calculating the LECR energy distribution at each location while solving the transport equation and then calculating the corresponding gamma-ray radiation spectrum. By iteratively solving these, we can ultimately obtain the coherent distribution of LECRs and gamma rays throughout space and time. This requires developing new numerical algorithms to improve computational efficiency.

\section{Summary and Outlook} \label{sec:summ}
This study proposes a novel method for investigating LECRs in protoplanetary disks via gamma-ray spectra. By solving the Parker transport equation, we simulated the propagation of LECRs in the protoplanetary disk and calculated the energy spectra of LECRs at different distances. Additionally, we simulated the nuclear reactions between LECRs and the gas and dust in the protoplanetary disk, producing nuclear de-excitation lines, and calculated the gamma-ray radiation spectra. This established a connection between the properties of LECRs and observable gamma-ray spectra. This method enables us to infer the composition, energy distribution, and spatial distribution of LECRs during the early stages of planetary system formation, thereby reconstructing the high-energy particle radiation environment of that period.

LECRs play an important role in various stages of planetary system formation. They can ionize and heat the gas in the protoplanetary disk, affecting chemical reactions and dust evolution, and thus impacting the formation and evolution of planetary systems. Therefore, accurately detecting and understanding the properties of LECRs is vital for understanding the mechanisms of planetary system formation. Traditional direct measurements face challenges, such as modulation by stellar wind, making our proposed method a valuable alternative for in-depth LECR research.

The future of LECR research is promising, with the emergence of new-generation MeV detector projects such as MeVGRO\endnote{\url{https://indico.icranet.org/event/1/} (accessed on 15 June 2024)} \cite{ref-yi2023}, MASS \cite{ref-zhu2024}, e-ASTROGAM \cite{ref-angelis2021}, AMEGO\endnote{\url{https://asd.gsfc.nasa.gov/amego/} (accessed on 15 June 2024)} \cite{ref-mcenery2019,ref-kierans2020}, COSI\endnote{\url{https://cosi.ssl.berkeley.edu/} (accessed on 15 June 2024)} \cite{ref-karwin2023}, and MeVCube \cite{ref-lucchetta2022}. With the deployment of these new high-resolution MeV gamma-ray telescopes, we are entering a golden age for LECR research. We note that these planned instruments mentioned above can hardly achieve an angular resolution better than $1^\circ$ in the MeV energy range, making the observation of protoplanetary systems difficult. Nevertheless, new techniques such as the Laue lens \cite{ref-ferro2023} can, in principle, provide an angular resolution of dozens of arcseconds, making the observation of the protoplanetary system in the MeV band feasible.

Through precise gamma-ray observations of protoplanetary disks, combined with advanced theoretical models, we will be able to reconstruct the real conditions of LECRs during the birth of planetary systems by obtaining information about the composition, energy spectrum, and spatial distribution of LECRs, thus understanding the high-energy particle environment of early planetary systems. We will also reveal the impact of CRs on planetary system formation by studying the effects of LECRs on chemical evolution, dust growth, and planet formation in protoplanetary disks, thereby understanding the origins of planetary system diversity. Furthermore, we will explore the CR environment of life's origin~\cite{ref-svensmark2006, ref-griessmeier2015, ref-rodgers2021} by assessing the impact of LECRs on the radiation environment and the origin of life on early Earth, and searching for habitable zones and clues to extraterrestrial~life.

In conclusion, utilizing gamma-ray detection of LECRs not only provides a unique approach to investigating the cosmic ray environment during the early stages of planetary system formation but also holds significant implications for deepening our understanding of the formation processes of stellar and planetary systems. This research will advance our comprehension of the universe, life, and the origins of humanity to new heights.

\vspace{6pt} 




\authorcontributions{Conceptualization, R.Y.; methodology, X.S.; software, X.S., Z.S., and B.L.; formal analysis, X.S.; investigation, X.S. and Z.S.; resources, R.Y.; data curation, X.S.; writing---original draft preparation, X.S.; writing---review and editing, Z.S., B.L., and X.S.; visualization, X.S. and S.Z.; supervision, R.Y.; project administration, R.Y.; funding acquisition, R.Y. All authors have read and agreed to the published version of the manuscript.}

\funding{Xulei Sun and Shuying Zheng acknowledge the support from the Undergraduate Innovative Training Program of the University of Science and Technology of China. Bing Liu acknowledges the support from the NSFC under the grant 12103049.}

\institutionalreview{Not applicable.}

\informedconsent{Not applicable.}

\dataavailability{The data underlying this article will be shared on reasonable request to the corresponding author.} 




\conflictsofinterest{The authors declare no conflicts of interest. The funders had no role in the design of the study; in the collection, analyses, or interpretation of data; in the writing of the manuscript; or in the decision to publish the results.} 


\newpage
\abbreviations{Abbreviations}{
The following abbreviations are used in this manuscript:\\

\noindent 
\begin{tabular}{@{}ll}
CRs & cosmic rays\\
LECRs & low-energy cosmic rays\\
LIS & local interstellar spectrum\\
ISM & interstellar medium
\end{tabular}
}

\appendixtitles{no} 
\appendixstart
\appendix
\section[\appendixname~\thesection]{} \label{AppA}
For the diffusion term in Equation \eqref{parker}
\begin{equation}
  \frac{\partial f}{\partial t}=\frac1{r^2}\frac\partial{\partial r}(r^2\kappa\frac{\partial f}{\partial r}),
\end{equation}
by introducing the variable substitution \(u = \ln r\), it turns into
\begin{equation}
  \frac{\partial f}{\partial t}=e^{-3u}\frac\partial{\partial u}(e^u\kappa\frac{\partial f}{\partial u}).
\end{equation}

Applying the Crank--Nicolson difference scheme to the above equation, we have
\begin{equation} \label{eq:cn-scheme}
  \begin{aligned}
    f_{i}^{n+1}=&\,f_i^n+e^{-3u_i}\frac{\Delta t}{2(\Delta u)^2}[e^{u_{i+1/2}}\kappa_{i+1/2}(f_{i+1}^{n+1}-f_i^{n+1})-e^{u_{i-1/2}}\kappa_{i-1/2}(f_i^{n+1}-f_{i-1}^{n+1})\\
    &+e^{u_{i+1/2}}\kappa_{i+1/2}(f_{i+1}^n-f_i^n)-e^{u_{i-1/2}}\kappa_{i-1/2}(f_i^n-f_{i-1}^n)].
  \end{aligned}
\end{equation}

Here, $f^{n}_i = f(t_n=n\Delta t, u_i = u_0 + i\Delta u)$, where $\Delta t$ and $\Delta u$ is the timestep and grid spacing on the $u$-axis, respectively, and $0 \le i \le L$.

By introducing
\begin{equation}
  \alpha_{i \pm 1/2} = \frac{\kappa_{i \pm 1/2}\Delta t}{(\Delta u)^2}e^{-3u_i+u_{i \pm 1/2}},
\end{equation}
the difference Equation \eqref{eq:cn-scheme} becomes
\begin{equation}
  \begin{aligned}
  2f^{n+1}_{i}=&\,2f^{n}_{i} + \alpha_{i+1/2}(f^{n+1}_{i+1} - f^{n+1}_{i}) - \alpha_{i-1/2}(f^{n+1}_{i} - f^{n+1}_{i-1}) \\
  &+ \alpha_{i+1/2}(f^{n}_{i+1} - f^{n}_{i}) - \alpha_{i-1/2}(f^{n}_{i} - f^{n}_{i-1}).
  \end{aligned}
\end{equation}

Rearranging the above equations, we have
\begin{equation}
  \label{eq:cn-scheme2}
  \begin{aligned}
    &-\alpha_{i+1/2}f^{n+1}_{i+1} + (2 + \alpha_{i+1/2} + \alpha_{i-1/2})f^{n+1}_{i} - \alpha_{i-1/2}f^{n+1}_{i-1} \\
    &= \alpha_{i+1/2}f^{n}_{i+1} + (2 - \alpha_{i+1/2} - \alpha_{i-1/2})f^{n}_{i} + \alpha_{i-1/2}f^{n}_{i-1}.
  \end{aligned}
\end{equation}

The linear system of Equation \eqref{eq:cn-scheme2} can be expressed in matrix form
\begin{equation}
  \mathsf{A}\mathsf{f}^{n+1} = \mathsf{B}\mathsf{f}^{n}+\mathsf{b},
\end{equation}
where the column vector $\mathsf{f}^{n} = [f^{n}_1, f^{n}_2, \cdots, f^{n}_{L-1}]^\top$, and the column vector $\mathsf{b} = [b_1 = \alpha_{1/2}(f^{n+1}_{0}+f^{n}_0), 0, \cdots, 0, b_{L-1}=\alpha_{L-1/2}(f^{n+1}_{L}+f^{n}_{L})]^\top$ is related to the boundary conditions. Here both $\mathsf{A} = [a_{mn}]$ and $\mathsf{B}=[b_{mn}]$ are $(L-1)\times(L-1)$ quasi-diagonal matrices,~where
\begin{adjustwidth}{-\extralength}{0cm}
\begin{equation}
  \begin{aligned}
   &a_{11} = -\alpha_{1/2},\ a_{12} = 2 + \alpha_{1/2} + \alpha_{3/2},\ a_{1k}=0\ (k \ne 0, 1),\\
   &a_{L-1,L-2} = -\alpha_{L-3/2},\ a_{L-1,L-1} = 2 + \alpha_{L-3/2} + \alpha_{L-1/2},\ a_{L-1,k}=0\ (k \ne L-2, L-1),
  \end{aligned}   
\end{equation}
\begin{equation}
  \begin{aligned}
   &b_{11} = \alpha_{1/2},\ b_{12} = 2 - \alpha_{1/2} - \alpha_{3/2},\ b_{1k}=0\ (k \ne 0, 1),\\
   &b_{L-1,L-2} = \alpha_{L-3/2},\ b_{L-1,L-1} = 2 - \alpha_{L-3/2} - \alpha_{L-1/2},\ b_{L-1,k}=0\ (k \ne L-2, L-1),
  \end{aligned}   
\end{equation}
\end{adjustwidth}
and for $1 < i < L-1$,
\begin{adjustwidth}{-\extralength}{0cm}
\begin{equation}
    a_{i,i-1} = -\alpha_{i-1/2},\ a_{i,i} = 2 + \alpha_{i+1/2} + \alpha_{1-1/2},\ 
    a_{i,i+1} = -\alpha_{i+1/2},\ a_{i,k} = 0 \ (k \ne i-1, i, i+1),
\end{equation}
\begin{equation}
    b_{i,i-1} = \alpha_{i-1/2},\  b_{i,i} = 2 - \alpha_{i+1/2} - \alpha_{1-1/2},\ 
    b_{i,i+1} = \alpha_{i+1/2},\ b_{i,k} = 0 \ (k \ne i-1, i, i+1).
\end{equation}
\end{adjustwidth}

\begin{adjustwidth}{-\extralength}{0cm}
\printendnotes[custom] 

\reftitle{References}

\PublishersNote{}
\end{adjustwidth}
\end{document}